# Quantum noise extraction from the interference of laser pulses in optical quantum random number generator

**ROMAN SHAKHOVOY**[1,2*], **DENIS SYCH**[1,2,3], **VIOLETTA SHAROGLAZOVA**[1,2,4], **ALEXANDER UDALTSOV**[1,2], **ALEKSEY FEDOROV**[1,2,5], AND **YURY KUROCHKIN**[1,2,6]

[1]*Russian Quantum Center, 45 Skolkovskoye shosse, Moscow, Russian Federation*
[2]*QRate, 100 Novaya str., Skolkovo, Russian Federation*
[3]*P.N. Lebedev Physical Institute, Russian Academy of Sciences, 53 Leninsky prosp., Moscow, Russian Federation*
[4]*Skolkovo Institute of Science and Technology, Bolshoy Boulevard 30, bld. 1, Moscow, Russian Federation*
[5]*Moscow Institute of Physics and Technology, 9 Institutskiy per., Dolgoprudny, Russian Federation*
[6]*NTI Center for Quantum Communications, National University of Science and Technology MISiS, 4 Leninsky prospekt, Moscow, Russian Federation*
*\*r.shakhovoy@goqrate.com*

**Abstract:** We propose a method for quantum noise extraction from the interference of laser pulses with random phase. Our technique is based on the calculation of a parameter, which we called the quantum reduction factor, and which allows determining the contributions of quantum and classical noises in the assumption that classical fluctuations exhibit Gaussian distribution. To the best of our knowledge, the concept of the quantum reduction factor is introduced for the first time. We use such an approach to implement the post-processing-free optical quantum random number generator with the random bit generation rate of 2 Gbps.

## 1. Introduction

Numerous quantum random number generators (QRNGs) based on various quantum effects have been demonstrated over the last two decades [1]. Among them, QRNGs employing different phenomena of quantum optics seem to be very convenient, relatively cheap, and, what is more important nowadays, could provide high random bit generation rates. Although optical QRNGs are extensively studied by many authors, the problem related to the contribution of classical noise in such QRNGs is still far from complete. This problem may seem overly pedantic at first glance. Perhaps for some scientific applications, such as Monte-Carlo simulations, it is. In cryptography, however, the extraction of purely quantum noise is crucial for secrecy and consequently is of fundamental importance.

Various approaches to evaluate the ratio between quantum and classical noises in optical QRNGs have been developed. Thus, in [2], where vacuum fluctuations were amplified using homodyne detection, the quantum noise contribution was estimated by calculating the difference between Shannon entropies of the amplified vacuum signal and the photodetector's dark signal. The SHA512 hash-function was then employed to extract quantum randomness from the raw bit sequence. A similar procedure for quantum noise extraction (but with other hash-function) was carried out in [3], where the interference of laser pulses with a random phase was proposed to generate random bits. For such a QRNG scheme, the same authors used later another approach [4], where the min-entropy of a random signal instead of the Shannon entropy was employed in the randomness extraction procedure. The same method to estimate the min-entropy was recently used in the optical QRNG of other authors [5], where the interference of pulses from a couple of gain-switched lasers was detected with balanced photodetector. To extract quantum randomness, the authors used then Toeplitz extractor [6, 7]. A more sophisticated approach of quantum noise extraction was used in [8, 9], where random bits were generated by using the interference of a continuous-wave laser radiation in a Mach-Zehnder interferometer. Authors used the fact that according to [10, 11] the quantum noise in their scheme is inversely proportional to the output laser power $P$. Recoding the dependence of the photodetector's voltage variance on $P$, they evaluated an optimal output laser power corresponding to the highest value of the quantum-to-classical noise ratio. The latter was then used to estimate the quantum min-entropy of the random signal needed for subsequent hashing.

An interesting method to make the correlated raw sequence uniform was proposed in [12], where authors used a finite-impulse response filter (FIR) to process the QRNG raw output. Such a filtering fuses bits of differing significances, thus achieving decorrelation of the raw data. Note, however, that such a processing is insecure in a cryptographic sense, since one can restore the raw sequence with all its inherent correlations, if the coefficients of the FIR filter are available. So, additional post-processing (i.e. randomness extraction) should be concatenated to the FIR, if such a QRNG is intended to use in cryptographic applications.

Another approach for data processing was proposed in [13], where a QRNG based on the interference of laser pulses was considered in the context of loophole-free Bell tests [14]. Authors considered the case of partially random bits in the output sequence and proposed the real-time method to increase its randomness. For this, neighboring bits were added modulo 2 recursively, i.e. the XOR operation was applied to the output sequence itself. The authors, however, noted that in case of completely predictable bits such an approach cannot be used. It is shown below that within our approach those bits that are related to the contribution of classical noise are (potentially) completely predictable; therefore, such a method is not applicable.

Obviously, there is no single rule to extract quantum noise from the output of the optical QRNG, particularly because the probability distribution of a random signal is highly dependent on the optical scheme. The quantum randomness extraction performed in [8] (and discussed later in detail by these authors in [7]) seems to be the most advanced approach to estimate quantum-to-classical noise ratio; however, this method is valid only when the interference term of the optical signal can be expanded into a series of the phase difference $\Delta\Phi$. In other words, such an approach can be applied only when the total phase fluctuations measured by the interferometric system are much less than unity, i.e., it is suitable only for the interferometers with sufficiently small time delay between the two arms [8]. Therefore, such method cannot be applied for the schemes with the interference of laser pulses with random phase [3-5], where $\Delta\Phi$ does not generally meet the requirement $\Delta\Phi \ll 1$. An attempt to extract quantum noise from the laser pulse interference was made in [4] (the same method was employed later in [5]); however, this approach seems to us not fully faithful, since it takes into account only the non-uniformity of the probability density function of the random signal, whereas the contribution of classical noise is not really taken into account. Moreover, it is not clear how to expand the proposed method for the case when a comparator is used to digitize a random signal instead of an analog-to-digital converter (ADC).

In the present work, we propose a different approach of the quantum randomness extraction for the optical QRNG based on the interference of laser pulses. Moreover, we propose a method to extract quantum noise without post-processing. In the next section, we provide some definitions that will be used across the paper. In section 3, we discuss main features of the interference signal and its probability distribution. In section 4, we discuss our method and introduce the so-called quantum reduction factor, which underlies the proposed approach. Finally, in section 5, we describe in detail the implementation of our QRNG.

## 2. Quantum vs classical randomness

Before discussing the problem of quantum randomness extraction, it is necessary to clarify what will we mean by randomness. Unfortunately, there is no widely accepted definition of random numbers. Just recall numerous definitions of random sequences given by D. Knuth [15] to realize the uncertainty surrounding this problem. Nevertheless, without giving definite

conclusion, D. Knuth gives a recipe stating that a useful definition should contain a short list of properties desirable for random sequences. We will follow this recipe and formulate the basic requirements for random sequences in practical applications of our interest, namely in quantum key distribution (QKD).

The first requirement is that the random sequence should be *nondeterministic*. Obviously, pseudorandom number generators producing numerical sequences that appear "random", does not satisfy this requirement, since they represent computer algorithms, which are deterministic by definition. Consequently, this requirement forces the use of physical entropy sources. However, not any physical entropy source can be considered nondeterministic. Thus, fluctuations of the gas pressure are generally considered to be a stochastic process essentially because of its collective nature: it is almost impossible to predict an exact value of the gas pressure in every moment of time just because it is extremely difficult to solve differential equations and substitute in the general solution the initial conditions for the velocities and coordinates of all the particles of the gas. Nevertheless, such fluctuations are fundamentally deterministic in a sense that it is possible in principle to solve corresponding differential equations and find a precise value of the gas pressure in any moment of time. We will refer such fundamentally deterministic entropy sources to as *classical*. In contrast, electron tunneling through a potential barrier or spontaneous emission of an atom are fundamentally nondeterministic processes. In fact, one cannot find out the exact time when the electron tunnels through a barrier or when an atom spontaneously emits a photon. There is only a finite probability that after the measurement we will obtain a given result. We will refer the entropy sources based on such phenomena to as *quantum*, and only quantum entropy sources will be treated as nondeterministic.

The second requirement is that the entropy source should be fundamentally *uncontrollable* by the third party. This just means that the QRNG should be designed in such a way that an adversary was not able to influence the result of measurements made in the systems. If it is impossible to exclude completely an impact of an adversary, his influence should be taken into account, for instance, by the use of postprocessing.

The third requirement is that the physical process used in the QRNG should be *unpredictable*. It might seem at first glance that quantum nondeterministic phenomena are automatically unpredictable, and this requirement is redundant. However, the result of a quantum mechanical measurement is not necessarily unpredictable in the general case. For example, polarization measurements with a pair of entangled photons are 100% correlated, if both polarizers are aligned along the same axis. That is, each photon may be found randomly either in channel (+) or (–) of the corresponding polarizer, but when photon 1 is found positively polarized, then its twin companion 2 is also found positively polarized [16]. Such quantum correlations can potentially be used by an adversary to find out a secret key; therefore, by the source of quantum entropy we will hereinafter mean a system, in which there are no quantum correlations available for measurement by the third party.

Summarizing the above, we will refer the nondeterministic, uncontrollable and unpredictable noise to as *quantum noise*. The term *classical noise*, in turn, will be used with respect to fluctuations, which are fundamentally deterministic in nature and could be controlled or predicted by the third party.

Finally, let us make a remark regarding the term "truly random". In cryptography, one usually deals with uniformly distributed random bit sequences, so "truly random" usually means here "uniformly distributed" [17, 18]. However, most physical entropy sources used to generate random signals do not always allow directly obtaining uniformly distributed random bit sequences, since signal fluctuations are rarely exhibit a uniform probability density function (PDF). Therefore, in the context of physical RNGs, "truly random" usually means "not pseudorandom" regardless the form of its distribution. In the framework of the noise classification given above, we will use below the term "truly random" only with regard to quantum noise. Moreover, to satisfy cryptography requirements, we will assume that random bit sequences generated by a QRNG are also uniformly distributed. So, under "truly random" we will understand both "quantum" and "uniformly distributed".

## 3. Probability density function of the interference signal

The optical scheme of the QRNG under consideration allows transforming the laser phase fluctuations into amplitude fluctuations. For this, a continuous sequence of laser pulses is entered into an unbalanced interferometer, whose delay line is selected such that the corresponding delay time is a multiple of the pulse repetition period, so that pulses emitted by the laser at different moments of time are met at the output of the interferometer. An important requirement for the operation of this scheme is that the laser should be modulated over the lasing threshold, i.e., after each pulse the laser should be switched to the amplified spontaneous emission (ASE) mode [19, 20]. Since most transitions in the ASE mode are spontaneous, phase correlations of the electromagnetic field are destroyed very quickly. As a result, each new laser pulse appears with a random phase; therefore, the result of the interference of two laser pulses will be a random quantity.

Let us make some remarks on the phase randomness in spontaneous emission process. It is well-known that spontaneous transitions are induced by zero-point oscillations of the electromagnetic field [21, 22]. ASE could be thus treated as amplified vacuum fluctuations; therefore, some authors [3, 4] use the relationship between vacuum fluctuations and spontaneous emission in order to attribute to latter the properties of vacuum, which is usually considered to be perfectly white, uncorrelated, and broadband. However, one should be careful when making such a generalization. In fact, perfect vacuum exhibit continuous set of states, whereas spontaneous emission in a laser is confined in its resonator with a finite number of modes, which changes the probability of spontaneous transitions [23]. Although individual spontaneous transitions are uncorrelated, correlation could exist between the phases of spontaneous emissions in a multilevel systems [24, 25]. Fortunately, in semiconductor laser spontaneous emission is only correlated for a carrier scattering time that is of order $10^{-13}$ s, a negligibly short time; therefore, spontaneous emissions can be considered to obey Markovian assumption [11]. Due to this (and not just because of the relation to vacuum fluctuations), one can treat the laser phase noise as quantum noise.

Let us now turn to the question of the interference of laser pulses. Here, we will neglect the contribution of relaxation oscillations into the pulse shape and will assume that it exhibits the Gaussian temporal profile. Moreover, we assume that the light in the interfering pulses has the same polarization. With these assumptions, the integral intensity $\tilde{S}$ of the interference signal can be written as follows:

$$\tilde{S} = s_1 + s_2 + 2\eta\sqrt{s_1 s_2}\cos\Delta\Phi, \tag{1}$$

where $s_1$ and $s_2$ are normalized integral intensities corresponding to the optical output from the short and long arms of the interferometer, respectively, $\eta$ is the visibility, and the phase difference is $\Delta\Phi = \Delta\varphi_p + \Delta\theta$. The phase difference $\Delta\theta$ is determined by the delay line $\Delta L$ and can be written as $\Delta\theta = k\Delta L n \omega_0/c$, where $n$ is the refractive index of the optical fiber, $\omega_0$ is the central frequency of the laser radiation, and the factor $k = 1, 2$ depends on the type of the interferometer (obviously, $k = 1$ if the Mach-Zehnder interferometer is used and $k = 2$ for the Michelson interferometer, since in this case the pulses pass the delay line twice). The phase difference $\Delta\varphi_p = \varphi_{p2} - \varphi_{p1}$, in turn, is determined by the initial phases of optical pulses at the laser output.

As discussed above, the phase of an optical pulse emitted by a gain-switched laser is assumed to be random. It should be noted here that such an assumption imposes some restrictions on laser operation, particularly on the pulse repetition rate $\omega_p$ and on the pump current amplitude $I_p$. In fact, at high $\omega_p$ the light coherence in the ASE mode could be destroyed incompletely [3, 4], such that the phase of subsequent laser pulses will correlate. The gain-switched laser will require more and more pump current when increasing the modulation frequency, which makes high demands on a current pulse driver. Moreover, negative correlation can occur between laser intensity fluctuations for weak excitations [26]. So, one should select the optimal values of the pump current and the pulse repetition frequency to make the interference random. As an example, in [4] satisfactory randomness was achieved for $\omega_p/2\pi = 5.825$ GHz with the reverse-biased distributed feedback (DFB) laser at $I_p \sim 100$ mA. Below, we assume that all parameters of the laser operation are set so that the randomness of the pulse phase is not disturbed. Particularly, the injection current of a laser is assumed to be always modulated over threshold with the modulation current amplitude not less than the threshold value. The pulse repetition period will be assumed to be always $\omega_p/2\pi = 2.5$ GHz.

It was shown that phase fluctuations in the ASE mode are well described by the Langevin equations in terms of phase diffusion [11]. Langevin forces driving phase fluctuations can be shown to be nearly Gaussian, such that random phases of laser pulses $\varphi_p$ can be assumed to be distributed according to the normal law with an rms of $\sigma_\varphi$. Obviously, the phase difference $\Delta\varphi_p$ between the two different laser pulses also has a normal PDF with an rms to be $\sigma_\varphi\sqrt{2}$. The same applies to the resulting phase difference $\Delta\Phi$ in Eq. (1). It can be shown (see Appendix) that if $\sigma_\varphi\sqrt{2} > 2\pi$ the PDF of the resulting phase $\Delta\Phi$ can be defined with high accuracy by

$$f_{\Delta\Phi} = \begin{cases} \dfrac{1}{\pi}, \Delta\Phi \in [0, \pi) \\ 0, \ \Delta\Phi \notin [0, \pi) \end{cases}. \qquad (2)$$

It should be noted that $\Delta\Phi$ could fluctuate under the influence of both quantum and classical noises. However, due to the fact that $\Delta\Phi$ is in the argument of the cosine (Eq. (1)), the influence of the classical component will be completely overlapped by quantum noise, if the rms of the quantum noise component is greater than $2\pi$. Indeed, the PDF of $\Delta\Phi$ in this case can be considered uniform within $[0, \pi)$ regardless the amount of the classical noise component. Hereinafter, we assume that the rms of the laser phase diffusion obeys the inequality $\sqrt{2}\sigma_\varphi > 2\pi$, such that $f_{\Delta\Phi}$ can be defined by Eq. (2) and fluctuations of $\Delta\Phi$ can be thus treated as truly random.

If $\Delta\Phi$ is distributed according to Eq. (2), whereas $s_1$, $s_2$ and $\eta$ are assumed to be constant, then the PDF of the integral signal $\tilde{S}$ is defined by the derivative: $f_{\tilde{S}}^Q = \left(F_{\tilde{S}}^Q\right)'$, where, by definition, the cumulative distribution function (CDF) $F_{\tilde{S}}^Q$ is given by [27]:

$$F_{\tilde{S}}^Q(y) = \int_{\tilde{S}<y} f_{\Delta\Phi}(x)dx, \qquad (3)$$

where $x$ stands for the value of $\Delta\Phi$ and the integration region is given by the inequality $s_1 + s_2 + 2\kappa\sqrt{s_1 s_2}\cos x < y$. Substituting Eq. (2) into Eq. (3) we obtain

$$f_{\tilde{S}}^Q(x) = \left[\pi\sqrt{(x-\tilde{S}_{min})(\tilde{S}_{max}-x)}\right]^{-1}, \qquad (4)$$

where

$$\tilde{S}_{min} = s_1 + s_2 - 2\eta\sqrt{s_1 s_2},$$
$$\tilde{S}_{max} = s_1 + s_2 + 2\eta\sqrt{s_1 s_2}. \quad (5)$$

We will refer the function $f_{\tilde{S}}^Q(x)$ given by Eq. (4) to as a *quantum* PDF of the interference signal, since it is defined solely by fluctuations of $\Delta\Phi$, which we agreed to consider quantum. The function $f_{\tilde{S}}^Q(x)$ for the case $s_1 = s_2 = 1$ at different values of visibility $\eta$ is shown in Fig. 1(a). One can see that $f_{\tilde{S}}^Q(x)$ tends asymptotically to infinity for ideal destructive ($x = \tilde{S}_{min}$) and constructive ($x = \tilde{S}_{max}$) interference. The "distance" between the asymptotes

$$\tilde{S}_{max} - \tilde{S}_{min} \equiv w_{\Delta\Phi} = 4\eta\sqrt{s_1 s_2}, \quad (6)$$

we will refer to as the width of the quantum distribution. One can see from Fig. 1 and Eq. (6) that $w_{\Delta\Phi}$ is decreased when decreasing $\eta$.

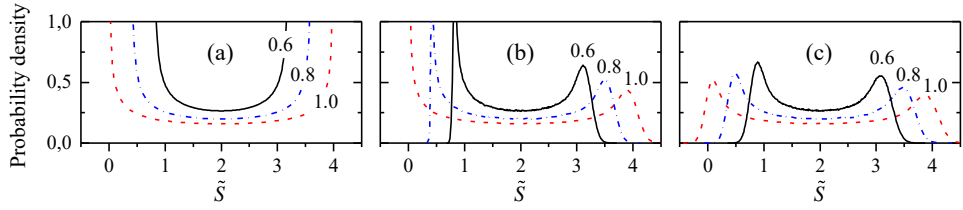

Fig. 1. (a) Quantum PDF of the interference signal (Eq. (4)) for three different values of the visibility $\eta$ (0.6, 0.8, and 1). (b) Monte-Carlo simulations of the signal PDF in the presence of fluctuation of $s_1$ and $s_2$ in Eq. (1). (c) Monte-Carlo simulations of the signal PDF in the presence of fluctuation of $s_1$ and $s_2$ and the photodetector's noise as well.

In addition to fluctuations of $\Delta\Phi$ one should take into account fluctuations of $s_1$ and $s_2$. The CDF of the interference signal should be then rewritten as follows

$$F_{\tilde{S}}(y) = \int_{\tilde{S}<y} f_{\Delta\Phi}(x_1) f_{s_1}(x_2) f_{s_2}(x_3) dx_1 dx_2 dx_3, \quad (7)$$

where the values of random variables $\Delta\Phi$, $s_1$ and $s_2$ are denoted by $x_1$, $x_2$, and $x_3$, respectively, and where it is assumed that fluctuations of $\Delta\Phi$, $s_1$ and $s_2$ are independent, such that the joint PDF represents just a product of all corresponding PDFs: $f_{\Delta\Phi} f_{s_1} f_{s_2}$. The integration area is defined now by the inequality $x_2 + x_3 + 2\eta\sqrt{x_2 x_3} \cos x_1 < y$. Note also that fluctuations of $\eta$ are neglected in Eq. (7). Finally, the resulting PDF of the interference signal is determined by a derivative of the CDF: $f_{\tilde{S}} = F'_{\tilde{S}}$. Unfortunately, the integral in Eq. (7) cannot be calculated analytically; therefore, Monte Carlo simulations are usually used to find $f_{\tilde{S}}$.

It seems reasonable to consider fluctuations of $s_1$ and $s_2$ as a Gaussian noise. Since this noise is related to the pump current fluctuations, it should be referred to as classical. Monte-Carlo simulations for the case when $f_{s_1}$ and $f_{s_2}$ are Gaussian with $\bar{s}_1 = \bar{s}_2 = 1$, whereas $f_{\Delta\Phi}$ is defined by Eq. (2), are shown in Fig. 1(b). One can see from the figure that the PDF exhibits noticeable asymmetry: the left maximum is much higher and "thinner" than the right one. This feature is due to fluctuations of normalized amplitudes $s_1$ and $s_2$ and it becomes more pronounced when increasing the rms value of these fluctuations. Note that the normalized rms value of the output laser power fluctuations $\sigma_s$ was usually measured to be 4-6% of the pulse average power, so the value $\sigma_{s_1} = \sigma_{s_2} = 0.05$ was used in simulations shown in Figs. 1(b,c).

In a real experiment, the PDF of the interference signal is additionally "broadened" due to noises in the photodetector. An experimental signal should be thus written in the following form:

$$\tilde{S} \to \tilde{S} + \zeta, \qquad (8)$$

where $\zeta$ is the photodetector's Gaussian classical noise. Simulations of $f_{\tilde{S}}$ in the presence of fluctuations of $s_1$, $s_2$ and the photodetector's noise as well (the rms of the photodetector noise was put to $\sigma_\zeta = 0.1$) are shown in Fig. 1(c).

It should be noted that the laser pulse interference could have another features, which adversely affect the visibility and have an impact on the appearance of the PDF of the random interference signal. Thus, we did not yet consider the influence of chirp and jitter, whose combined effect in the context of QRNG was considered in [12], where authors demonstrated that the PDF of the interference signal for chirped laser pulses differs markedly from the PDF measured in the absence of chirp. It is well-known that Gaussian laser pulses exhibit linear chirp [19], such that the time dependence of the electric field of the pulse is proportional to $\exp[i(\omega_0 t - \beta t^2)]$, where $\omega_0$ is the central frequency of the laser field and the linear chirp coefficient is $\beta = \alpha/2w^2$, where $w$ is the rms width of the laser pulse and $\alpha$ is the linewidth enhancement factor (the Henry factor [10]). The visibility of the integrated interference signal in Eq. (1) is now defined by [28]

$$\eta = e^{-\frac{(1+\alpha^2)\Delta t^2}{8w^2}}, \qquad (9)$$

where $\Delta t$ is the inaccuracy of pulse overlap, which fluctuates due to jitter. Therefore, Eq. (7) should be supplemented by the jitter PDF $f_{\Delta t}$, which is usually assumed to be Gaussian.

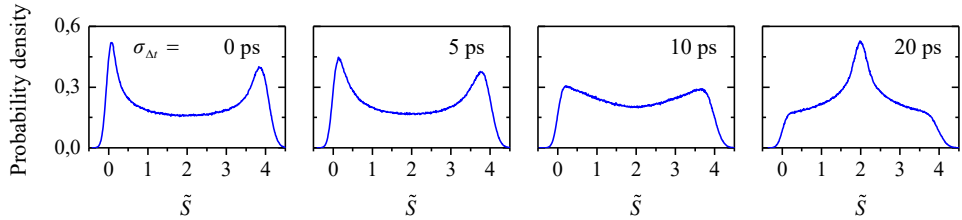

Fig. 2. Monte-Carlo simulations of the signal PDF taking into account the influence of the "linear chirp + jitter" effect. The values of the jitter rms are shown on the corresponding simulations.

Monte-Carlo simulations of $f_{\tilde{S}}$ taking into account the influence of the "linear chirp + jitter" effect are shown in Fig. 2. For simulations, we used Eq. (1) with the visibility $\eta$ defined by Eq. (9), where we put $\alpha = 6$ and $w = 50$ ps. Fluctuations of $s_1$ and $s_2$ were again assumed to be Gaussian with $\bar{s}_1 = \bar{s}_2 = 1$ and $\sigma_{s_1} = \sigma_{s_2} = 0.05$, $f_{\Delta\Phi}$ was defined by Eq. (2) and the photodetector noise was introduced according to Eq. (8) with $\sigma_\zeta = 0.1$. The jitter was assumed to exhibit Gaussian PDF with the rms from 0 to 20 ps and with mean value equal to zero. One can see that the jitter with $\sigma_{\Delta t} > 10$ ps markedly affects the form of the signal PDF leading to the appearance of the central peak, which indicates an increase in the probability that the signal equals to $\tilde{S} = s_1 + s_2$, which is the evidence of interference worsening. However, at small $\sigma_{\Delta t}$ (or, equivalently, at small $\alpha$) the influence of the "linear chirp + jitter" effect on the PDF is insignificant (compare PDFs on Fig. 2 at $\sigma_{\Delta t} = 0$ and $\sigma_{\Delta t} = 5$ ps), such that one can neglect it. Moreover, the "linear chirp + jitter" effect can be reduced by cutting off the high-frequency and low-frequency parts of the laser spectrum with the bandpass filter. This is the consequence of the fact that the spectrum of the chirped Gaussian laser pulse exhibits inhomogeneous broadening in addition to broadening associated with a finite pulse duration. The spectral filtering changes the intensity distribution of spectral components in the pulse making it effectively less chirped.

A more complicated picture takes place when the laser pulse is distorted by relaxation oscillations, since the chirp is not linear in this case. A more detailed study of this issue, particularly the influence of chirp, jitter and relaxation oscillations on laser pulse interference we consider elsewhere [29], showing that the combined effect of chirp and jitter can be decreased even in this case by cutting off only the high-frequency part of the laser spectrum with the bandpass filter. Thus, we will assume below that the "chirp + jitter" effect is decreased to be small enough, such that the visibility $\eta$ is not significantly changed from one pair of pulses to another and we can treat it as a non-fluctuating parameter.

The key to separate quantum and classical noises is the comparison of functions $f_{\tilde{S}}$ and $f_{\tilde{S}}^Q$. Obviously, the more these functions are different from each other, the greater the contribution of classical fluctuations. In the next section, we describe a method to calculate the so-called quantum reduction factor, which in a sense determines the quantum-to-classical noise ratio in the assumption of Gaussian classical noises. As the main sources of classical noise, we will consider the photodetector noise and fluctuations of the laser output power (the latter correspond to fluctuations of quantities $s_1$ and $s_2$). In view of the above, fluctuations of the visibility $\eta$ and corresponding impact of jitter will be neglected. We will further restrict our consideration to the case when an adversary can influence only the part of the classical noise, which is related to the photodetector. It means that if the magnitude of the classical noise in the QRNG system varies with time, this variation can be mainly attributed to the photodetector and a possible influence of an attacker on it. This does not mean, however, that under this assumption we exclude from consideration another part of the classical noise related to fluctuations of the laser output power. The latter will still be included in the model by Eq. (7) and thus will make its impact into the quantum reduction factor. The only assumption is that the magnitude of this part of the classical noise, i.e. the values of $\sigma_{s_1}, \sigma_{s_2}$, will be assumed to be fixed.

## 4. Quantum reduction factor

As we mentioned in the Introduction, quantum and classical noises can be "separated" at the post-processing stage. In fact, it can be formally assumed that the output random sequence may contain correlations associated not only with non-uniformity of the digitized signal, but also with the contribution from the classical noise. This means that the randomness extraction (RE) procedure should be carried out taking into account the ratio between quantum and classical noises.

The RE procedure can be considered as some operation that transforms a binary sequence $\{0,1\}^l$ of length $l$ with *non-uniform* distribution of elements into a binary sequence $\{0,1\}^m$ of length $m$, where distribution of elements is *close to uniform*. The length of the binary sequence with improved randomness is generally shorter: $m < l$. With such a definition, the RE procedure can be treated as a reduction of a raw random bit sequence:

$\{0,1\}^l \xrightarrow{\text{RE}} \{0,1\}^m$, and a ratio $\gamma = l/m$ is sometimes referred to as a reduction factor. In conventional RE procedures, the reduction factor $\gamma$ is estimated via the min-entropy $H_\infty$ of the raw sequence. Thus, a perfect randomness extractor applied to a non-uniform random sequence $\{X_1, X_2, ..., X_N\}$ with $N \gg 1$, where each $X_i$ is an $n$-bit word, could provide $NH_\infty$ almost uniformly distributed bits [17], i.e. the raw sequence will be reduced with such an extractor by a factor of $\gamma = nN/NH_\infty = n/H_\infty$. The min-entropy, in turn, is defined as $H_\infty = -\log_2 p_{max}$, where $p_{max}$ is the highest probability to guess a random element from the sequence $\{X_1, X_2, ..., X_N\}$. If the random signal is digitized by an ADC, then $n$ in the definition of $\gamma$ corresponds to the resolution of the ADC in bits, whereas $p_{max}$ corresponds to the probability of the most likely bin. If the digitization is performed using a comparator ($n = 1$), then the reduction factor is determined as $\gamma = 1/H_\infty$. If, in addition, the comparator threshold is chosen such that probabilities of '1's and '0's in the QRNG's output are equal, then $H_\infty = 1$ and the reduction factor is $\gamma = 1$, i.e., a raw random sequence could be employed. Obviously, such a result contradicts physical considerations, since the classical noise introduced by the photodetector and other devices included in the QRNG cannot be generally neglected, so the raw random bit sequence should be subject to reduction anyway. Therefore, the reduction factor should be redefined to take into account classical noises.

It seems that there is no universal way to estimate contributions of classical and quantum noises to laser pulse interference. One can see from Fig. 2 that the "chirp + jitter" effect complicates the appearance of the signal PDF and it is not obvious how to compare functions $f_{\tilde{S}}$ and $f_{\tilde{S}}^Q$, when $f_{\tilde{S}}$ exhibit significant central peak (see Fig. 2 on the right). As we already agreed, we will neglect the influence of the "chirp + jitter" effect on the signal PDF and will consider the photodetector noise and fluctuations of the laser output power as the main sources of the classical noise. Assuming further that the interference signal is digitized with the comparator, we can quite easily estimate the contribution of classical fluctuations.

One can see from Fig. 1 that the Gaussian noise broadens the PDF of the interference signal, such that the probability for the signal to fall in the region between $\tilde{S}_{min}$ and $\tilde{S}_{max}$ decreases when increasing the rms of the photodetector noise. We can thus say that an additional classical entropy "flows" into the $[\tilde{S}_{min}, \tilde{S}_{max}]$ interval. Let us agree that if the contributions of classical and quantum noises are equal in this interval, we will not trust the resulting random sequence at all (even if it passes all randomness tests!) and require the reduction factor to be made infinitely large. (We will discuss such an assumption below.) In contrast, if the contribution of classical noises is negligibly small, then the reduction factor can be put to unity (note that this assumption is valid only for the case of a comparator with a properly chosen threshold). Such a reduction factor that takes into account the contribution of classical noise and allows extracting pure quantum randomness we will refer to as a *quantum reduction factor* $\Gamma$. Let us now find the relation between $\Gamma$ and the min-entropy.

In the ideal case, when the classical contribution is absent, the comparator threshold voltage (or rather its normalized value) should be obviously set to $V_{th} = \tilde{S}_{min} + w_{\Delta\Phi}/2$, and the min-entropy can be written as follows:

$$H_\infty^Q = -\log_2 \left( \int_{\tilde{S}_{min}}^{\tilde{S}_{min} + w_{\Delta\Phi}/2} f_{\tilde{S}}^Q(x)dx \right) = 1, \tag{10}$$

where the integral in parentheses corresponds obviously to $p_{max}$. We will refer $H_\infty^Q$ to as a quantum min-entropy. The min-entropy in the presence of classical noise we define in a similar way:

$$H_\infty = -\log_2\left(\int_{\tilde{S}_{min}}^{\tilde{S}_{min}+w_{\Delta\Phi}/2} f_{\tilde{S}}(x)dx\right) \geq 1. \quad (11)$$

Following the above agreement, we will assume that if $H_\infty$ is twice $H_\infty^Q$, then $\Gamma \to \infty$. If, however, $H_\infty \to H_\infty^Q$, then $\Gamma \to \gamma = 1/H_\infty$. Obviously, both requirements are satisfied, if the quantum reduction factor is defined as follows:

$$\Gamma = \frac{1}{2-H_\infty}. \quad (12)$$

It is obvious from the above that $H_\infty \geq H_\infty^Q$ and, consequently, $\Gamma \geq \gamma$, and the equality holds in the absence of classical noises. Thereby, the reduction factor $\gamma$ determines the non-uniformity degree of a random sequence, but it does not take into account the contribution of classical noise. The quantum reduction factor $\Gamma$, in turn, takes into account both effects and thus allows estimating the length of the random bit sequence returned by the RE algorithm, which will be guaranteed to have a quantum nature.

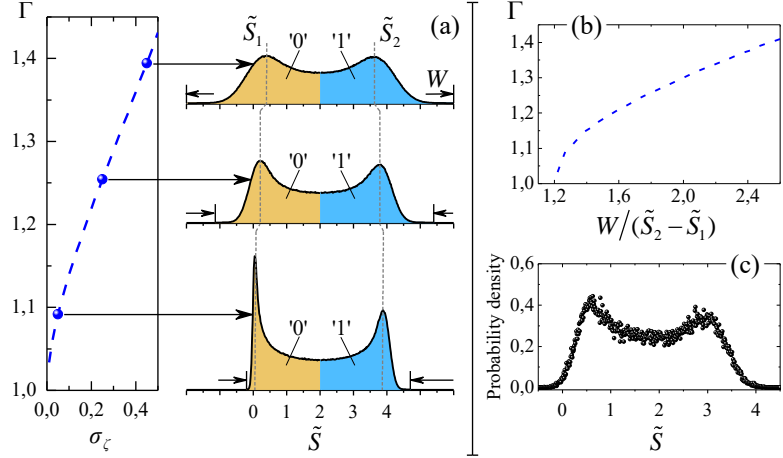

Fig. 3. (a) The theoretical dependence of the quantum reduction factor $\Gamma$ on the photodetector noise width $\sigma_\zeta$ in case of digitization of the interference signal by the comparator. Monte-Carlo simulations of the PDF of the interference signal $\tilde{S}$ corresponding to the three different values of $\sigma_\zeta$ are shown to the right of the curve. (b) The theoretical dependence of the quantum reduction factor $\Gamma$ on the PDF broadening factor $B = W/(\tilde{S}_2 - \tilde{S}_1)$. (c) Experimental PDF of the interference signal.

The theoretical dependence of the quantum reduction factor $\Gamma$ on the photodetector noise rms $\sigma_\zeta$ is shown in Fig. 3(a) on the left. The simulations of the integral interference signal $\tilde{S}$ PDF corresponding to three different values of $\sigma_\zeta$ are shown on the right. It was assumed in the simulations that the photodetector noise is included in $\tilde{S}$ according to Eq. (8); the fluctuations of $s_1$ and $s_2$ were again assumed to be Gaussian with $\bar{s}_1 = \bar{s}_2 = 1$ and $\sigma_{s_1} = \sigma_{s_2} = 0.05$. The selected points on the curve $\Gamma(\sigma_\zeta)$ are connected by arrows with the corresponding theoretical PDFs. One can see that $\Gamma$ grows with the growth of $\sigma_\zeta$, since the proportion of the noise, which may be compromised by the adversary, increases.

One should remember that $\Gamma$ depends also on $\sigma_{s_1}$ and $\sigma_{s_2}$, which are assumed to be fixed in the present consideration. Assuming that $\sigma_{s_1} = \sigma_{s_2} = \sigma_s$ we may write for a more general case $\Gamma = \Gamma(\sigma_\zeta, \sigma_s)$, such that the quantum reduction factor can be presented as a 2D surface or as a set of $\Gamma(\sigma_\zeta)$ curves corresponding to different magnitudes of laser pulse fluctuations. We will not consider this case here.

There is a certain arbitrariness in the definition of the quantum reduction factor given by Eq. (12). In fact, we demand that $\Gamma$ should be put to infinity when the contributions of classical and quantum noises are the same. Probably, such a requirement is overly rigid, but it guarantees that the random sequence resulting from the RE procedure with such a reduction factor will indeed have a quantum nature. We use the min-entropy as a measure to compare quantum and classical noises not only because $H_\infty$ is used in the definition of $\gamma$, but also because such a choice seems very natural. Indeed, we do not trust the noise, if the probability for the signal $\tilde{S}$ to fall into the interval from $\tilde{S}_{min}$ to $w_{\Delta\Theta}/2$ is changed from 1/2 to 1/4, i.e. when the min-entropy doubles. In this case, the probability of a '0' or '1' is equally related to both quantum and classical effects, i.e. quantum and classical noises become in a sense indistinguishable.

This interpretation can be expanded for the case $n > 1$, i.e. when an ADC is used for the digitization. Let us again require the quantum reduction factor to become infinitely large when the probability $p_{max}$ of the most likely bin is halved due to classical noise contribution. We define this probability now as follows:

$$p_{max} = \int_{\tilde{S}_{min}}^{\tilde{S}_{min}+\Delta u} f_{\tilde{S}}^Q(x) dx, \quad (13)$$

with the bin size $\Delta u = \Delta U/2^n$, where $\Delta U$ is the dynamic range of the ADC, and where we use the fact that $f_{\tilde{S}}^Q(x)$ behaves asymptotically near $\tilde{S}_{min}$. The quantum min-entropy is obviously defined as $H_\infty^Q = -\log_2 p_{max}$, whereas the value of the min-entropy, at which $\Gamma \to \infty$, is $-\log_2(p_{max}/2) = 1 + H_\infty^Q$. So, we can define the quantum reduction factor as follows:

$$\Gamma = \frac{n}{1 + H_\infty^Q - H_\infty}, \quad (14)$$

where similar to Eq. (11)

$$H_\infty = -\log_2 \left( \int_{\tilde{S}_{min}}^{\tilde{S}_{min}+\Delta u} f_{\tilde{S}}(x) dx \right) \quad (15)$$

and $H_\infty^Q$ is defined accordingly, but with $f_{\tilde{S}}^Q$ in the integral. One can see that with such a definition Eq. (12) becomes an extreme case of Eq. (14) at $n = 1$, if $\Delta u$ is treated as $w_{\Delta\Phi}/2$.

## 5. QRNG implementation

The schematic diagram of our QRNG is shown in Fig. 4(a). The optical scheme depicted by the dashed rectangle includes generally two principal elements: the fiber optic interferometer (it is an unbalanced Michelson interferometer in our case) and the photodetector. Note that the optical scheme in Fig. 4(a) may generally refer to any scheme that allows implementing the interference of laser pulses. Optical pulses are generated by the distributed feedback laser modulated over threshold with the frequency 2.5 GHz by a laser diode driver. To digitize the photodetector signal we propose to use the set of three high-speed comparators. The comparator C0 is needed to find the PDF of the interference signal, whereas the comparators C1 and C2 work in parallel acquiring the signal from the photodetector and providing the digital output. Obtained random bits are received by the field-programmable gate array (FPGA) for buffering and further processing.

To determine the signal PDF, we propose to sweep the comparator C0 threshold voltage, $V_{th}^0$, recording a random bit sequence of a specified length for each value of $V_{th}^0$ and calculating then the corresponding ratio of ones and zeroes in the current sequence: $R = N_{ones}/N_{zeroes}$. One can then restore the value of the signal PDF corresponding to the $i$-th value of $V_{th}^0$ using the following relation:

$$f_{\tilde{S}}^i = \frac{|R_i - R_{i+1}|}{\Delta V \left(1 + R_i + R_{i+1} + R_i R_{i+1}\right)}, \tag{16}$$

where $\Delta V$ is the voltage sweep step. Note that throughout the article by photodetector or comparator voltage we mean dimensionless quantity related to the normalized signal $\tilde{S}$ and not to the signal in volts.

Generally, only a single comparator, C1 or C2, is needed to obtain a random output, so let us assume for now that only one of them is used. The purpose of the second comparator will be clarified shortly. By definition, if the photodetector signal exceeds the threshold voltage of the comparator, the latter outputs a logical one, otherwise the signal from the comparator corresponds to a logical zero. The threshold voltage should be chosen so that the ratio of the number of ones to the number of zeros in the output random sequence was close to unity. Since we know the signal PDF $f_{\tilde{S}}$ found with the comparator C0, we can calculate the threshold voltage by defining it such that the areas under $f_{\tilde{S}}$ left and right of the threshold were equal.

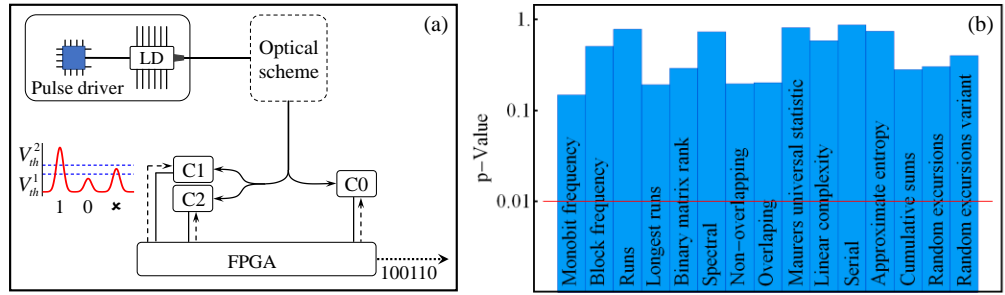

Fig. 4. (a) The schematic diagram of the QRNG: C0, C1, and C2 are high-speed comparators, LD is the laser diode, $V_{th}^1$ and $V_{th}^2$ in the inset stand for the threshold voltages of the comparators C1 and C2, respectively. (b) The result of the NIST statistical suite for one of the obtained raw random bit sequences. To pass the test we imposed the condition: p-Value $\geq 0.01$.

Using an arrangement with a single working comparator, we acquired the 1Mbit random sequences. The data were then extracted from the FPGA buffer and stored as binary files on the PC. All of them successfully passed all NIST tests [30]. The result of the NIST statistical suite for one of the obtained sequences is shown in Fig. 4(b).

As mentioned above, the raw random bit sequences cannot be employed despite the successful randomness tests, since the raw signal is "diluted" by the classical noise. So, these sequences should be subject to randomness extraction procedure and thus the quantum reduction factor should be calculated. Unfortunately, the formulas for $\Gamma$ given above cannot be applied directly, since the calculation of $H_\infty$ with Eq. (11) requires the knowledge of $\tilde{S}_{min}$, which defines the integration limits. Obviously, one cannot calculate $\tilde{S}_{min}$ knowing only $f_{\tilde{S}}$; therefore, other approach should be used.

One of the possible methods is the substitution of $\tilde{S}_{min}$ by the value of the normalized integral signal $\tilde{S}_1$ corresponding to the left maximum of the PDF. In fact, one can see from Fig. 3(a) that $\tilde{S}_1$ is quite close to the left asymptote of $f_{\tilde{S}}^Q$ (i.e. to $\tilde{S}_{min}$), if the classical Gaussian noise is quite small. However, this maximum shifts to the right when increasing the classical noise contribution, which obviously overestimates the value of $\Gamma$. In fact, this method applied to the experimental PDF shown in Fig. 3(c) provides $\Gamma = 2.23$, which is unreasonably high.

Alternatively, one can use the fact that the PDF becomes broader when increasing the classical noise. For definiteness, we will assume that the width $W$ of the distribution $f_{\tilde{S}}$ corresponds to the range, where $f_{\tilde{S}} > 10^{-5}$ (see Fig. 3(a)). Note that this choice is arbitrary, and with the same success one could take, e.g., $f_{\tilde{S}} > 10^{-4}$; however, this will obviously change abscissa values in Fig. 3(b). Therefore, it is important to choose the same assignation for both theoretical and experimental PDFs when calculating $W$. The "distance" between its maxima, in turn, decreases, so we can introduce the dimensionless quantity $B = W/(\tilde{S}_2 - \tilde{S}_1)$, which reflects the contribution of classical noises. Here, $\tilde{S}_1$ and $\tilde{S}_2$ stand for the values of the integral signal, corresponding to the left and right maximum of $f_{\tilde{S}}$, respectively (see Fig. 3(a)). The dependence $\Gamma(\sigma_\varsigma)$ can be thus substituted by the dependence $\Gamma(B)$, which is shown in Fig. 3(b). The main advantage of this representation is that $B$ does not depend explicitly on $H_\infty$ and can be easily calculated from the experimental PDF. The value of $\Gamma$ can be then easily found from Fig. 3(b). We estimated the experimental value of the broadening factor to be $B = 1.77$, which provides $\Gamma = 1.25$. Comparing the PDF shown in Fig. 3(a) in the middle (it corresponds to $\Gamma \approx 1.25$) with the experimental one shown in Fig. 3(c), it becomes obvious that this estimate is more reasonable. So, after the RE procedure (we used hashing) the raw random sequence is reduced by a factor 1.25 resulting in the random bit generation rate of 2 Gbps.

Finally, let us consider the role of the comparators C1 and C2 and show how the quantum noise can be extracted without post-processing. Note that if the photodetector signal falls near the center of the PDF, i.e. if the photodetector output is close to the comparator threshold, then there is a high probability that the resulting bit is forged by an intruder. In fact, in this case an adversary could "toss" the output left and right of the threshold using his influence on the classical noise. Therefore, one should discard signals corresponding to some region near $V_{th}$ in order to avoid intrusion of an adversary. The width of such a region should be guaranteed to be larger than the width of classical fluctuations. Discarding untrusted bits is analogous to the reduction of the output sequence, so the width of the "untrusted region" should be related to the value of the quantum reduction factor $\Gamma$. Denoting the area under $f_{\tilde{S}}$ between the boundaries of the untrusted region as $P$ and taking into account that the remaining area is $1 - P$, we can define the quantum reduction factor as follows: $\Gamma = 1/(1-P)$. The value of $P$, in turn, can be defined by the integral

$$P = \int_{V_{th}-\Delta V_\Gamma^1}^{V_{th}+\Delta V_\Gamma^2} f_{\tilde{S}}(x)dx, \qquad (17)$$

where $\Delta V_\Gamma^1$ and $\Delta V_\Gamma^2$ are "untrusted intervals" left and right of the threshold. One can see from Fig. 3 that the PDF is quite symmetric in the vicinity of $V_{th}$, so one can put $\Delta V_\Gamma^1 = \Delta V_\Gamma^2 = \Delta V_\Gamma$ and using the definition of $\Gamma$ in terms of $P$ write the following relation:

$$\int_{V_{th}}^{V_{th}+\Delta V_\Gamma} f_{\tilde{S}}(x)dx = \frac{\Gamma-1}{2\Gamma}, \qquad (18)$$

which defines the interval $\Delta V_\Gamma$.

Afterall, the threshold voltages of the comparators C1 and C2 are set to $V_{th}^1 = V_{th} - \Delta V_\Gamma$ and $V_{th}^2 = V_{th} + \Delta V_\Gamma$, respectively. The digital output from the two comparators should be then added modulo 2. Let us denote the output of the comparators C1 and C2 as $c_1$ and $c_2$, respectively. If $c_1 \oplus c_2 = 0$, then the FPGA buffers $c_1$ or $c_2$ (either one of them, since they are the same in this case). If, however, $c_1 \oplus c_2 = 1$, then nothing is written to the buffer (see the inset in Fig. 3(a)). So, discarding the signal that falls into the range from $V_{th} - \Delta V_\Gamma$ to $V_{th} + \Delta V_\Gamma$, we improve the reliability of the random bit sequence. This method can be thus considered as a hardware quantum randomness extractor.

Let us summarize the working process of the QRNG presented in Fig. 4(a). We assume first that the laser continuously generates short pulses at 2.5GHz repetition rate. The working cycle of the QRNG starts with the calculation of $f_{\tilde{S}}$ with the comparator C0 using Eq. (16). For this, one should specify the step $\Delta V$ of the threshold voltage sweep and the number of bits that will be used to find ratio of ones and zeroes at each value of $V_{th}^0$. Calculated density distribution is then saved as an array in the memory. Then the threshold $V_{th}$ is calculated such that the areas under $f_{\tilde{S}}$ left and right of $V_{th}$ were equal. Then the PDF broadening factor $B$ is calculated and the quantum reduction factor $\Gamma$ is determined from the theoretical dependence $\Gamma(B)$. Knowing $\Gamma$ and $V_{th}$ the system calculates $\Delta V_\Gamma$ with Eq. (18) and sets threshold voltages for the comparators C1 and C2. In parallel, the system again starts calculating $f_{\tilde{S}}$, $V_{th}$ and $\Gamma$ performing thus the on-the-fly control of the QRNG operation. Afterwards, the FPGA starts buffering random bits checking for each sample the result of the XOR operation of the digital signals from the comparators and discarding the samples for which $c_1 \oplus c_2 = 1$.

Note that the embodiment of the QRNG with a single working comparator, where the post-processing is employed, is somewhat equivalent to the implementation with the two comparators C1 and C2, where the hardware quantum randomness extraction is performed. However, due to its simplicity, the latter seems to us more preferable. Note also that the raw random bit sequences were already "random enough" to pass the statistical tests, so processed sequences obviously pass them too; therefore, we do not present the results of the tests here.

It is important to mention that despite the ideological similarity between the post-processing procedure and the hardware quantum randomness extraction presented here, they do different jobs. Conventional randomness extractors assume that the pseudorandom sequence is somehow correlated, and these correlations are removed with the use of, e.g., cryptographic hash-functions, which transform the raw sequence such that it becomes unrecognizable. The quantum randomness extraction procedure developed here assumes that quantum noise is truly random by default, but the whole noise of the system is contaminated by classical noise, which could be (albeit not necessarily) correlated. In contrast to hashing, the proposed hardware quantum noise "extractor" does not remove correlations from the bit sequence but eliminates the contribution of classical noise. Acquired random bit sequence is then assumed to be related to the pure quantum noise and thus considered to be perfectly random. If we compare the raw sequence and the sequence obtained after our extractor, they will be very similar, with the only difference being that some bits in the "pure quantum" sequence will be skipped. Therefore, it is not quite correct to compare the hardware "extractor" reported here with conventional randomness extractors developed for pseudo-random numbers in terms of latency, usability or speed. The only common feature between them is the measure of reduction of the raw sequence length, which is defined by the quantum reduction factor $\Gamma$. Note also, that the proposed hardware quantum noise extractor was defined only for the scheme with the comparator. For the scheme with an ADC, the method described above is not applicable, since one cannot just "cut off" the center of the signal PDF in this case.

**Conclusions**

We demonstrated a simple method of quantum noise extraction from the interference of laser pulses. The developed approach is based on the calculation of the quantum reduction factor $\Gamma$, which allows determining the contributions of quantum and classical noises in the assumption that classical fluctuations exhibit Gaussian PDF. To the best of our knowledge, the concept of the quantum reduction factor is introduced for the first time. It was shown how to calculate $\Gamma$ for the case, when an ADC is used to digitize the signal, as well as for the case when the comparator is used for the digitization.

A robust scheme of the QRNG with the random bit generation rate of 2 Gbps was proposed. We developed a method for the on-the-fly control of the QRNG operation based on the continuous calculation of the signal PDF followed by the hardware randomness extraction. Due to its simplicity, the proposed randomness extraction procedure seems to be a good alternative to conventional post-processing procedures employing cryptographic hash-functions, Toeplitz extractors, etc.

**Appendix**

As we mentioned in the main text, semiconductor laser phase fluctuations are well described by the Langevin equations in terms of phase diffusion [11]. The random phases of laser pulses $\varphi_p$ can be assumed to be distributed according to the normal law with an rms of $\sigma_\varphi$, whereas the phase difference $\Delta\varphi_p$ between the two different laser pulses also has a normal PDF with an rms to be $\sigma_{\Delta\Phi} = \sigma_\varphi\sqrt{2}$. The PDF of the phase difference $\Delta\Phi = \Delta\varphi_p + \Delta\theta$ may be then written in the following form

$$f_{\Delta\Phi}(x) = \frac{1}{\sigma_{\Delta\Phi}\sqrt{2\pi}} \exp\left(-\frac{(x-\Delta\theta)^2}{2\sigma_{\Delta\Phi}^2}\right). \tag{19}$$

Since $\Delta\Phi$ is in the argument of the cosine (Eq. (1) in the main text), then taking into account that the value of $\cos(\Delta\Phi)$ will not change neither after the substitution $\Delta\Phi \to \Delta\Phi + 2\pi j$ ($j$ is integer) nor after the change of the sign $\Delta\Phi \to -\Delta\Phi$, we can write $f_{\Delta\Phi}$ as follows:

$$f_{\Delta\Phi}(x) \leftrightarrow \begin{cases} \sum_{p=\pm 1}\sum_{j=-\infty}^{\infty} f_{\Delta\Phi}(px+2\pi j), x \in [0,\pi) \\ 0, \; x \notin [0,\pi) \end{cases}, \tag{20}$$

whence

$$f_{\Delta\Phi}(x) = \frac{J\left(\frac{x}{2} - \frac{\Delta\theta}{2}, e^{-\sigma_{\Delta\Phi}^2/2}\right) + J\left(\frac{x}{2} + \frac{\Delta\theta}{2}, e^{-\sigma_{\Delta\Phi}^2/2}\right)}{2\pi}, \qquad (21)$$

where $J(u,q)$ is the Jacobi theta function:

$$J(u,q) = 1 + 2\sum_{j=1}^{\infty} q^{j^2} \cos(2ju). \qquad (22)$$

Since in our case $q < 1$, the series in (22) rapidly converges, so the value of the theta function can be estimated with the use of just the two first terms:

$$J(u,q) = 1 + 2q\cos 2u. \qquad (23)$$

It is obvious from Eq. (23) that the deviation of $J(u,q)$ from unity is determined by the factor $2q = 2\exp(-\sigma_{\Delta\Phi}^2/2)$. Already at $\sigma_{\Delta\Phi}^2 = (2\pi)^2$ we have $2q \sim 10^{-8}$, so one can assume with great accuracy that

$$f_{\Delta\Phi} = \begin{cases} \frac{1}{\pi}, & \Delta\Phi \in [0,\pi) \\ 0, & \Delta\Phi \notin [0,\pi) \end{cases}, \qquad (24)$$

if $\sigma_{\Delta\Phi} = \sigma_\varphi \sqrt{2} > 2\pi$.

## Funding

Russian Science Foundation (Grant No. 17-71-20146)

## Disclosures

The authors declare no conflicts of interest